# Graphene separation and stretching induced by piezoelectric effect of ferroelectric domains: impact on the conductance of graphene channel

Anna N. Morozovska[1,2], Anatolii I. Kurchak[3], and Maksym V. Strikha[3,4*]

[1] *Institute of Physics, National Academy of Sciences of Ukraine,*
*Prospect Nauky 46, 03028 Kyiv, Ukraine*

[2] *Bogolyubov Institute for Theoretical Physics, National Academy of Sciences of Ukraine,*
*14-b Metrolohichna street, 03680 Kyiv, Ukraine*

[3] *V. Lashkariov Institute of Semiconductor Physics, National Academy of Sciences of Ukraine,*
*Prospect Nauky 41, 03028 Kyiv, Ukraine,*

[4] *Taras Shevchenko Kyiv National University, Radiophysical Faculty*
*Prospect Akademika Hlushkova 4g, 03022 Kyiv, Ukraine.*

**Abstract**

P-N junctions in graphene on ferroelectric have been actively studied, but the impact of piezoelectric effect in ferroelectric substrate with ferroelectric domain walls (FDWs) on graphene characteristics was not considered. Due to the piezo-effect ferroelectric domain stripes with opposite spontaneous polarizations elongate or contract depending on the polarity of voltage applied to the substrate. We show that the alternating piezoelectric displacement of the ferroelectric domain surfaces can lead to the alternate stretching and separation of graphene areas at the steps between elongated and contracted domains. Graphene separation at FDWs induced by piezo-effect can cause unusual effects. In particular, the conductance of graphene channel in a field effect transistor increases essentially, because electrons in the stretched section scatter on acoustic phonons. At the same time the graphene conductance is determined by ferroelectric spontaneous polarization and varies greatly in the presence of FDWs. The revealed piezo-mechanism of graphene conductance control is promising for next generations of graphene-based field effect transistors, modulators, electrical transducers and piezo-resistive elements. Also our results propose the method of suspended graphene fabrication based on piezo-effect in a ferroelectric substrate that does not require any additional technological procedures.

---

[*] Corresponding author. E-mail: maksym.strikha@gmail.com

# I. INTRODUCTION

Experimental and theoretical studies of remarkable electromechanical, electrophysical and transport properties of graphene remain on the top of the researchers attention since graphene discovery [1, 2] till nowadays [3, 4, 5]. Despite prominent advances in understanding of the complex physical processes taking place in graphene and other 2D-semiconductors, these materials are not commercially used in modern electronic techniques despite very attractive application possibilities. Most challenges associated with the practical usage of these materials critically depend on the successful choice of electromechanical, electrophysical and physicochemical properties of their environment (substrates, matrices, or gates). At that the environment should be compatible with a given 2D-material and desirable to have additional functionality degrees [3, 6].

A promising and quite feasible way towards the commercialization of graphene-based devices (as well as the devices utilizing other 2D semiconductors) is to use "smart" substrates with additional (electromechanical, polar and/or magnetic) degrees of functionality, which are coupled with graphene. Of particular interest is a graphene on a ferroelectric substrate [7, 8, 9, 10, 11,], whose spontaneous polarization and domain structure can be controlled by an external electric field [7-6, 12, 13]. For example, the polar state of ferroelectric substrate can be readily switched by the voltage applied to the gate of the graphene field effect transistor (**GFET**) [7, 6], where a graphene or other 2D-semiconductor sheet is a channel.

At that the presence of a domain structure in a ferroelectric substrate can lead to the formation of p-n junctions in graphene [12, 13]. The junctions are located near the contact of the domain walls with the ferroelectric surface [14, 15, 16]. Note that the unique properties of the p-n junction in graphene have been realized much earlier by multiple gates doping of graphene channel by electrons or holes, respectively [17, 18, 19]. Then they have been studied theoretically [20, 21] and experimentally [22, 23, 24]. However, only relatively recently Hinnefeld et al [12] and Baeumer et al [13] explored the advantages to create a p-n junction in graphene using ferroelectric substrates. Notably this way imposes graphene on a 180$^o$-ferroelectric domain wall (**FDW**). Due to the charge separation by an electric field of a FDW – surface junction [25, 26], p-n junction can occur without any additional gates doping. Later on semi-quantum and semi-phenomenological analytical models have been developed for different types of carrier transport (ballistic, diffusive, etc) in a single-layer graphene channel at 180$^o$-FDW [14, 15]. The dynamics of p-n junctions in graphene channel induced by FDW motion have been studied [16].

To the best of our knowledge, all existing theoretical models considering electrotransport in graphene on ferroelectric substrate with FDWs do not consider the impact of piezoelectric effect in the substrate on graphene strain (see e.g. [14-16, 20, 21, 25, 26]). However it is well-known that the elastic strain can change the band structure of graphene (e.g. via the deformation potential) and open the band gap [4, 5, 27, 28, 29]. As will be shown in this paper the elastic strain can significantly affect the graphene conductance via the **stretching** of its surface and **separation** of graphene areas at the steps between elongated and contracted domains.

The estimates made on the back of the envelope show that if the voltage $U$ is applied to a gate of the GFET with FDW, one domain elongates and another one contracts depending on the voltage polarity [compare **Fig. 1(a)** with **1(b)**]. Corresponding surface displacement can be significant for ferroelectrics with high piezoelectric coefficients. For instance, the piezoelectric coefficients of $PbZr_{0.5}Ti_{0.5}O_3$ (**PZT**) substrate can reach $(0.3 - 1)$nm/V depending on the film thickness and temperature [30]. Hence the piezoelectric effect in PZT leads to its surface displacement $h \sim (0.5 - 1)$ nm for the gate voltage $\sim(1 - 3)$V. The thickness $d \leq 0.5$ nm of the physical gap between the graphene and ferroelectric is determined by Van-der-Waals interaction. The density $J$ of binding energy for graphene on $SiO_2$ substrate is about 0.5 J/m$^2$ [31]. Because graphene adhesion to mica surface is considered to be the strongest one in comparison with other surfaces, it is natural to expect that $J$ value for graphene on PZT surface is smaller. On the contrary the Young's modulus of graphene is extremely high ($Y = 1$ TPa, [32, 33]).

Note that the height $h$ of ferroelectric surface displacement step induced by the piezoelectric effect at FDWs is usually much smaller than their width $w$ [see designations in **Fig. 1(b)**], since typically $h \leq 1$nm and $w \geq 5$ nm [34]. If the step width $w$ is small enough in comparison with the domain size $D$, we can say that the displacement of ferroelectric surface at FDW has **"sharp profile"**. Under these conditions the partially separated graphene region can occur at the step. This happens when the normal component $F_n$ of the elastic tension force $F$ applied to a carbon atom exceeds the force $F_b$ binding the atom with the surface [see the scheme of the forces in **Fig. 1(b)**]. The separated section of length $l + \Delta l$ is "**suspended**" between the bounded sections. Its horizontal projection has length $l$, at that the small difference $\Delta l \ll l$ can be neglected since the height $h \ll w$ and typically $l \geq w$ [see designations in **Fig. 1(b)**]. For observable graphene separation we require $D > (2 - 3)w$ (or even $D \gg w$).

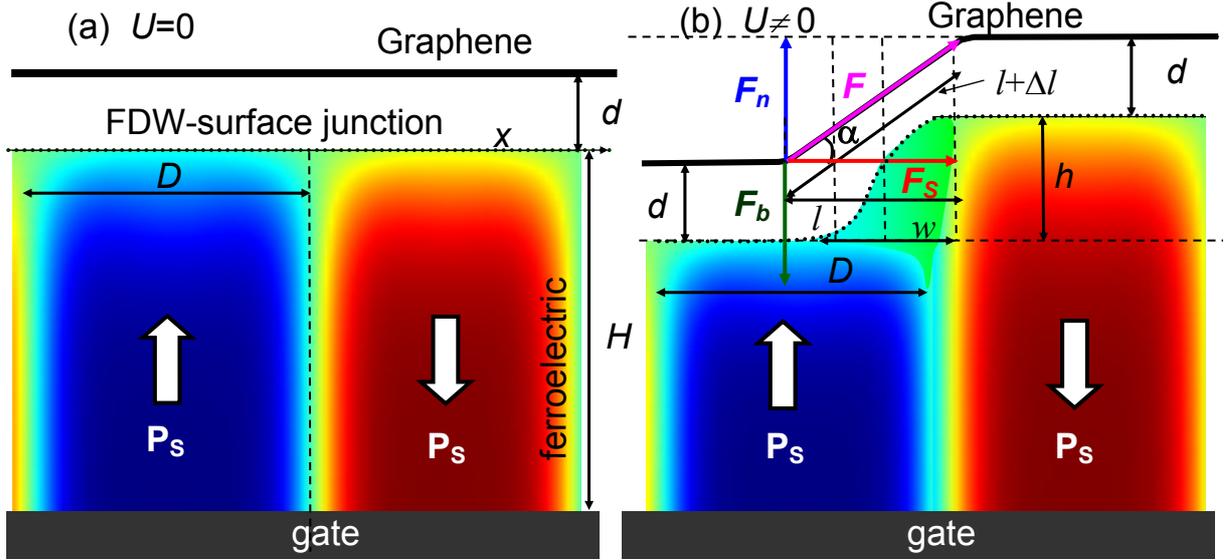

**FIG. 1.** Partial separation of graphene channel sections induced by a piezoelectric effect at the ferroelectric domain wall – surface junction. The separation is absent at *U*=0 **(a)** and appears at *U*≠0 **(b)**. In plot **(b)** *F* is the elastic tension force, $F_n$ is its normal component, $F_S$ is its lateral component and $F_b$ is the binding force of the carbon atom to the surface.

Therefore it does make sense to analyze more rigorously the piezoelectric displacement of ferroelectric surface in the vicinity of FDW-surface junction. The analysis is presented below.

## II. PIEZOELECTRIC DISPLACEMENT OF FERROELECTRIC SUBSTRATE SURFACE

Let us analyze the vertical displacement $u_3(x)$ of ferroelectric film surface in the vicinity of FDW-surface junction induced by the piezoelectric effect. The voltage *U* is applied to the gate electrode. Analytical expression for $u_3(x)$ is derived in Ref.[34] within the framework of decoupling approximation that is widely used for the piezoelectric response calculations [35, 36, 37, 38]. The displacement $u_3(x)$ acquires the form [34]:

$$u_3(x) = -U[W_{33}(x)d_{33} + W_{31}(x)d_{31}]. \tag{1}$$

Here *U* is the potential difference between the top and bottom electrodes, i.e. the gate voltage. Constants $d_{33}$ and $d_{31}$ are piezoelectric coefficients. They are temperature-dependent constants for thick ferroelectric films and tabulated for the bulk materials.

The presence of the first term proportional to $d_{33}$ in Eq.(1) is anticipative because the spontaneous strain in the bulk of ferroelectric with the spontaneous polarization $P_3$ is proportional to $d_{33}$. The origin of the second term proportional to $d_{31}$ in Eq.(1) depends on the mechanical boundary conditions between the ferroelectric film and the gate electrode. The concrete forms of $W_{33}(x)$ and $W_{31}(x)$ are dictated by the conditions of the film and substrate electrode mechanical compatibility. Note that the ferroelectric film and gate electrode have close mechanical properties in a realistic situation of perovskite-on-perovskite epitaxial growth (e.g. for the pair $PbZr_xTi_{1-x}O_3$ on metallic $SrRuO_3$). Assuming that the mechanical properties of the film and electrode are close, we can use the rigorous derivation of the terms in Eq.(1) presented in Ref. [34]. For the case the functions $W_{33}$ and $W_{31}$ are:

$$W_{33}(x) = \frac{2}{\pi}\left[\left(\frac{d}{H}+1\right)\arctan\left(\frac{x}{d+H}\right) - \left(\frac{d}{H}\right)\arctan\left(\frac{x}{d}\right)\right] + \frac{2}{\pi}\frac{x}{H}\ln\left(\frac{(d+H)^2+x^2}{d^2+x^2}\right) \quad (2)$$

$$W_{31}(x) = \frac{2(1+2\nu)}{\pi}\left[\left(\frac{d}{H}+1\right)\arctan\left(\frac{x}{d+H}\right) - \left(\frac{d}{H}\right)\arctan\left(\frac{x}{d}\right)\right] + \frac{\nu}{\pi}\frac{x}{H}\ln\left(\frac{(d+H)^2+x^2}{d^2+x^2}\right) \quad (3)$$

Here $H$ is the thickness of ferroelectric film, $d$ is the distance between the flat surface of ferroelectric and graphene, and $\nu$ is the Poisson ratio [39]. Since the distance $d$ is defined by the Van-der-Waals forces, it is typically small (not more than 1 nm) [3] and so Eqs.(1)-(3) can be simplified in the limit $d \to 0$ (see **Appendix A**).

Notably, finite size effects and imperfect screening conditions (such as physical gaps [40] between the film and graphene, and / or dead layers [41] near the ferroelectric surface) can strongly decrease the value of the spontaneous polarization in thin films. The effects eventually leads to the size-induced phase transition to a paraelectric phase with the film thickness decrease below the critical thickness $H_{cr}$. At that the critical thickness $H_{cr}$ depends on the gap and/or dead layer thickness in a self-consistent way, and usually does not exceed (5-15) nm for ultra-thin gaps (see e.g. Refs [42, 43]). As a result the piezoelectric coefficients $d_{33}$ and $d_{31}$ included in Eqs.(1),(3) and next equations are no more constants defined for the bulk material, but become polarization dependent. In the simplest case the changes can be estimated from the expression $d_{ijk} = 2\varepsilon_0\left(\varepsilon^f_{km} - \delta_{km}\right)Q_{ijml}P_l^S$, where $P_l^S$ are the components of spontaneous polarization, $\varepsilon^f_{km}$ are components of relative dielectric permittivity tensor and $Q_{ijml}$ are the components of electrostriction tensor. On the other hand the dead layer thickness $H_{DL}$ and graphene-surface separation $d$ contribute additively to the critical thickness, namely $H_{cr} \sim H_{DL} + d$ [44]. The self-consistent problem of $d_{ijk}$ determination becomes analytically tractable for ultra-thin gaps and dead layers (smaller than 1-2 lattice constants of ferroelectric and much smaller than the film thickness $H$), and enough thick films ($H \gg H_{cr}$) only. Basing on the numerical results presented in Refs.[16, 43], we can conclude that if the strong inequalities $H \gg H_{DL} + d$ and $H \gg H_{cr}$ are valid, one can regard $P_l^S$, $d_{33}$ and $d_{31}$ equal to their tabulated values for a bulk material.

**Figure 2** illustrates the profiles of the surface displacement calculated from Eqs.(1)-(2) at gate voltage $U$=1 V, thermodynamic piezoelectric coefficients $d_{33} \approx 10^3$ pm/V and $d_{31} \approx -450$ pm/V and Poisson ratio $\nu \approx 0.3$ corresponding to PZT at room temperature [30]. Values of the PZT film thickness $H$ and distance $d$ varies in the range $H = (20 - 500)$ nm and $d = (0 - 0.5)$ nm, respectively. Note that vertical picometer scale is much smaller than the horizontal nanometer scale in **Fig.2**.

One can see from the figure that the step originated at FDW-surface junction is the widest for the smallest ratio $d/H$ (curve 1 has the half-width of about 250 nm), and becomes essentially thinner with the ratio increase (curves 2-5 have the half-width of about (5 –50) nm). The dashed curve 5 calculated for $d = 0$ and $H = 100$ nm is almost indistinguishable from the curve 2 calculated for $d = 0.5$ nm and $H = 100$ nm. Hence the step width is not defined by the ratio $d/H$ only, it appeared proportional to both of these values. Actually, the full step width $w$, defined as the distance between its maximal positive and negative displacement (shown by a horizontal arrow in **Fig.2**), can be estimated from the expression:

$$w \approx k(H + d), \qquad (4)$$

where the numerical factor $k$ depends on the ratio $d/H$ and appears smaller than unity for the curves 2-5 shown in **Fig.2**.

The step maximal height $h$ changes in a non-monotonic way with $d/H$ ratio increase, but the displacement difference is the same for all curves far from the domain wall. Actually, far from the wall, $x \to \pm\infty$, functions $W_{3i}(x)$ in Eq. (2)-(3) tend to $\pm 1$. Hence, the "saturated" step height $h_\infty = |u_3(x \to \infty) - u_3(x \to -\infty)|$ is

$$h_\infty = 2|U|[d_{33} + (1 + 2\nu)d_{31}]. \qquad (5)$$

The value $h_\infty$ is independent on the ferroelectric thickness and proportional to the product of the gate voltage and combination of the piezoelectric coefficients $d_{33} + (1 + 2\nu)d_{31}$.

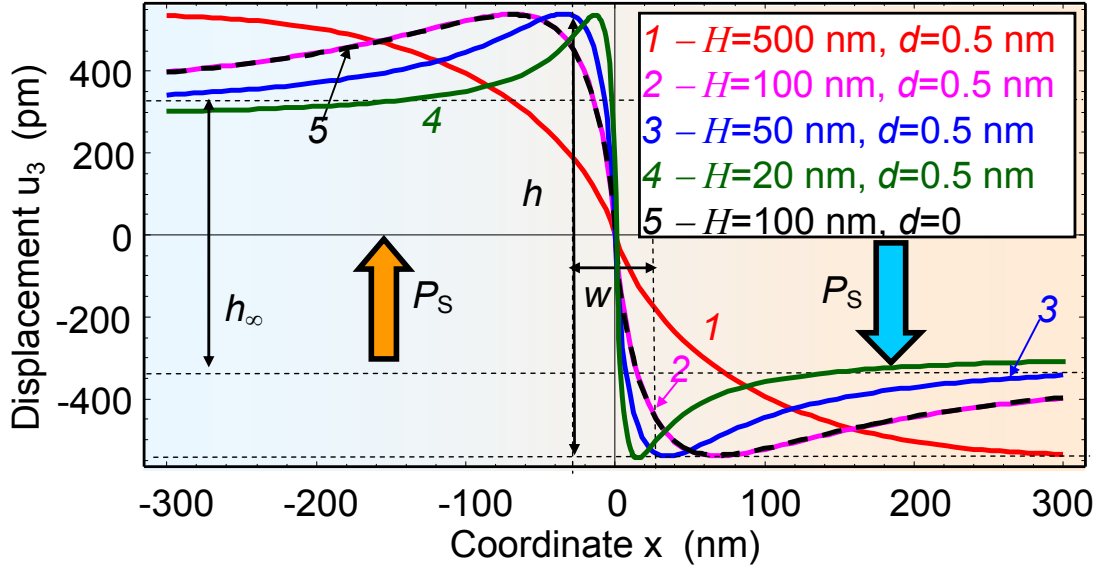

**FIG. 2.** Profiles of the ferroelectric surface displacement $u_3$ calculated from Eqs.(1)-(3) at the gate voltage U=1 V, thermodynamic piezoelectric coefficients $d_{33} \approx 10^3$ pm/V, $d_{31} \approx -450$ pm/V and Poisson ratio $v$=0.3 corresponding to PZT at room temperature. The values of ferroelectric film thickness $H$ and distance $d$ for the curves 1 - 4 are indicated in the legend. The step height $h$ and width $w$ for the curve 3 are shown by vertical and horizontal arrows respectively.

For the case $d \ll H$ the graphene stretching (without separation) can be favorable in the case of its contact with thick ferroelectric films characterized by smooth displacement profiles across the FDWs (see curve 1 in **Fig. 2**).

Note that the situation shown in **Fig.2** corresponds to an artificial case of a single domain wall in a ferroelectric film. In reality ferroelectric films (of thickness less than dozens of microns) inevitably split into stripe domains if their surfaces are not in a perfect electric contact with ideally conducting electrodes [45, 46]. The domain splitting occurs due to the long-range nature of the depolarization electric fields [46]. The incomplete surface screening of ferroelectric polarization strongly influences the domain nucleation and growth dynamics, domain walls structure and period in thin films under incomplete screening conditions [46]. The conditions are open-circuit electric boundary conditions [46, 47], imperfect electrodes [48], separation from the electrodes by ultra-thin dead layers [40] or spatial gaps [41]. Since the graphene layer is separated from the ferroelectric surface by the ultra-thin gap of thickness $d$ (see **Figs. 1** and **3**), the domain splitting can occur. So the question about the relationship between the domain period and film thickness should be considered. In the simplest Kittel-type models the period of domain stripes $D$ with infinitely thin walls depends on the film thickness $H$ in accordance with Kittel-Mitsui-Furuichi (**KMF**) relation [46, 49], $D = 2\sqrt{h_M H}$, where the length $h_M \cong A\varepsilon_0 \left(1 + \sqrt{\varepsilon_{11}\varepsilon_{33}}\right) \psi_{DW} / P_S^2$ depends on the concrete model and ferroelectric parameters. The numerical coefficient $A \geq 3.7$, surface energy of the FDW $\psi_{DW} \sim (0.1 - 0.5)\,\text{J/m}^2$, effective dielectric permittivity $\sqrt{\varepsilon_{11}\varepsilon_{33}} \sim (10^2 - 10^3)$ and spontaneous polarization $P_S \sim (0.05 - 0.5)\,\text{C/m}^2$ depend on the film thickness and temperature. Estimates show that $h_M$ appears about (1 – 10) nm for e.g. PbTiO$_3$ at room temperature [43]. Notably, that the KMF law $D \sim \sqrt{H}$ appears invalid for LGD-type models, which naturally accounts for FDWs broadening near ferroelectric surfaces separated by the gap from the conductor [43]. In numbers the domain period $D$ essentially increases with the screening degree increase reaching hundreds of nm for ultra-thin gap $d \sim 0.5$ nm used in this work. Therefore the strong inequality $D \gg w$ is possible even in relatively thin films. Another challenge for experimental situation, presented in **Figs.1** and **3**, is how to prevent the uncontrollable motion and splitting of the separated FDWs in thin films under the gate voltage increase. However, lattice potential and defects pin the walls rather strongly (see chapter 8, subsection 8.5 in Ref.[42] and refs therein, as well as Refs. [50, 51, 52, 53]).

### III. PIEZOELECTRIC EFFECT IMPACT ON GRAPHENE LAYER

The complete separation (i.e. exfoliation) of graphene caused by a piezoelectric effect is hardly possible in thick ferroelectric films with smooth profiles of the surface displacement ($w \sim D$) corresponding to the curve 1 in **Fig.2**. The complete exfoliation becomes impossible if the domain wall is a single one in a film; and so only the stretching of graphene sheet is induced by the piezoelectric effect in the case. The typical picture of graphene stretching by a piezoelectric effect in a thick film is shown in **Fig.3(a).**

The partial separation or complete exfoliation of graphene can be favorable when it contacts with relatively sharp ferroelectric surface profiles across the FDWs ($w \ll D$) corresponding to the curves 2-5 in **Fig.2**. A typical picture of graphene partial separation or complete exfoliation caused by a piezoelectric effect in a thin film is shown in **Fig.3(b)** by solid red and magenta curves, respectively**.** The separated sections are suspended and fixed between the bounded graphene regions. The stretching of the suspended graphene regions can be strong enough in the case of its partial fixing by ferroelectric surface [see red curve 1 in **Fig.3(b)**], if the Van-der-Waals forces can hold the graphene at the bottom of the cavities induced by piezoelectric effect on domain structure. Note that the graphene stretching is almost absent in the case of its complete exfoliation from ferroelectric surface due to the weakness of Van-der-Waals forces, and the graphene sheet is suspended above the bottoms of domains wells in the case [see magenta curve 2 in **Fig.3(b)**].

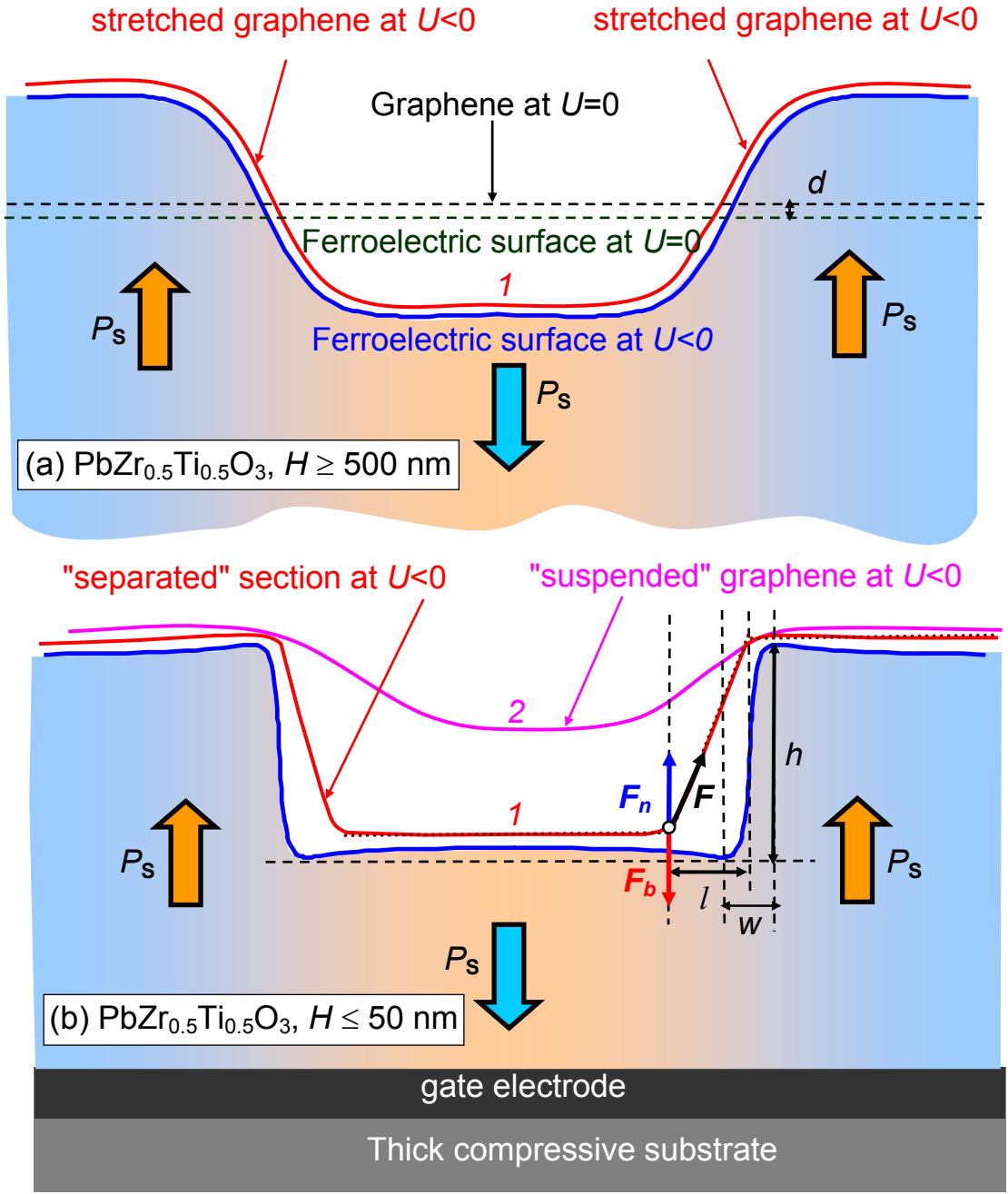

**FIG. 3.** Schematic profiles of the ferroelectric film and graphene surface displacements for $U=0$ (dashed horizontal lines) and $U < 0$ (solid curves 1, 2). **(a)** Graphene stretching (without any separation) induced by the piezoelectric effect in thick films with smooth profile of the surface displacement across the FDWs (red curve 1). **(b)** Partial (red curve 1) and complete (magenta curve 2) separation of graphene region induced by piezoelectric effect in thin films with sharp profile of the surface displacement across the FDWs. Note that vertical scale is much smaller than the horizontal one and so $l \gg h$ [see **Fig.2** for clarity].

Which case (stretching, partial separation or complete exfoliation) can be realized in a real system? Only self-consistent numerical calculations of the elastic sub-problem allowing for *ab initio* calculations of the binding energy $J$ and distance $d$ can give answer to the question. Unfortunately the calculations, being beyond the scope of this work, are absent to date. Hence all possible cases of graphene mechanical behavior can be realized for a given piezoelectric displacement of the ferroelectric film surface.

Thus depending on the ratios $D/H$ and $d/h$ the situation with graphene stretching and partial separation can be very different (see **Figs. 3**). Taking into account the warning, the length of the graphene section separated from the ferroelectric surface can be estimated for the sharp profile of the surface displacement, at that graphene surface profile is approximated by dotted lines as shown in **Fig.3(b)**. The limits for this model applicability will be discussed later. It is natural to expect that the graphene separation occurs right to the point, where the normal component $F_n$ of the tangential force $F$ and binding force $F_b$ are equal [see **Fig.1(b)** and **Fig.3(b)**]. Taking into account obvious expressions for the forces listed in **Appendix B**, after elementary calculations presented there, we derived analytical expression for the minimal length $l$ of separated graphene region,

$$l = h \cdot \sqrt[3]{\frac{Yd}{2J}} > |U|(d_{33} + (1+2\nu)d_{31})\sqrt[3]{\frac{4Yd}{J}}. \tag{6}$$

The inequality in Eq.(6) originates from the inequality $h > h_\infty$ [see **Fig.2** with designations]. Estimates made from Eq.(6) give that the stretched section can reach tens of nm for gate voltages $|U| \geq 3$ V, PZT parameters at room temperature, binding energies $J < 0.25$ J/m$^2$ and separation $d$=0.5 nm. and. For a suspended graphene the length can be much longer, namely $l \sim D$ [see magenta curve 2 in **Fig.3(b)**].

It is obvious now that the linear approximation of the separated graphene region, shown by a dotted line in **Fig.3(b)**, can be used for the case when the length $l$ of separated graphene region is at least longer than the halfwidth $w/2$ of the ferroelectric surface displacement step at the FDW. Substituting the estimate (6) and Eq.(4) at $d \ll H$ in the inequality $l > w/2$ we obtain the condition of the linear approximation validity:

$$|U|(d_{33} + (1+2\nu)d_{31})\sqrt{\frac{4Yd}{J}} > \frac{k}{2}H. \qquad (7)$$

For chosen PZT parameters the inequality (7) becomes valid for e.g. relatively thin films [$H < 50$ nm], gate voltages $U$ higher than 3.5 V, binding energies $J<0.25$ J/m$^2$ and graphene-ferroelectric separation $d=0.5$ nm.

Note that thin films (tens nm or thinner) of multiaxial ferroelectric Pb$_x$Zr$_{1-x}$TiO$_3$ with the composition x near the morphotropic boundary x=0.5 and without perfect electric contact between its surfaces and electrodes can change their polarization direction from the out-of-plane to in-plane one [54, 55]. In-plane or closure domains can appear with the film thickness decrease in order to minimize the depolarization field energy in the gap between graphene layer and ferroelectric surface. For the case the inequality (7) loses its sense. However it appeared that simple measures can be performed for Eq.(7) validity. Firstly, the in-plane polarization direction can be more stable than the out-of-plane one in thin PZT film only under the absence of misfit strain or for tensile strains [54, 55]. A compressive misfit strain $u_m$ about −0.01 or more can stabilize the out-of-plane polarization in PZT [54, 55, 56] and essentially decrease the critical thickness $H_{cr}$ of ferroelectricity disappearance [57], that can be estimated from the formulae,

$H_{cr} \approx \dfrac{H_{DL}+d}{\varepsilon_0\varepsilon_d\left[\alpha_T(T-T_C)-2Q_{12}u_m/(s_{11}+s_{12})\right]}$ [58], where $\alpha_T = 2.66 \times 10^5$ C$^{-2}$·J·m/K, $T_C \approx 666$ K is a ferroelectric Curie temperature, $T$ is the ambient temperature, $Q_{12} \approx -0.0295$ C$^{-2}$·m$^4$ is the negative electrostriction coefficient, $s_{11}=8.2\times10^{-12}$ Pa$^{-1}$ and $s_{12}= -2.6\times10^{-12}$ Pa$^{-1}$ are elastic compliances [54, 59], $\varepsilon_d$ is the relative dielectric permittivity of the gap and $\varepsilon_0 = 8.85\times10^{-12}$ F/m is the universal dielectric constant. Due to the orientating role of compressive substrate the critical thickness of multidomain film can become five lattice constants or even less [60]. Hence it is enough to deposit the epitaxial PZT film on a mechanically rigid thick compressive substrate, like perovskite SrTiO$_3$ [see **Fig.3(b)**].

The piezo-effect induced separated areas of graphene, which are "suspended" between elongated and shortened domains, can cause many interesting effects, some of which will be discussed below.

## IV. GRAPHENE CONDUCTANCE

The conductance of graphene channel in diffusion regime can change essentially, because electrons in the separated stretched section scatter on acoustic phonons [21]. In particular the voltage dependence of the conductance $G(U)$ of graphene channel, when its part of length $l(U)$ is separated and another part of length $L - l(U)$ is bounded, obeys the Matiessen rule [21]:

$$G(U) = W \left[ \frac{L - \chi|U|}{\sigma_B} + \frac{\chi|U|}{\sigma_S} \right]^{-1} \qquad (8)$$

Here $L$ is the channel length and $W$ is its width. The separated length $l(U) = \chi|U|$, where the coefficient $\chi = (d_{33} + (1 + 2\nu)d_{31})\sqrt[3]{4Yd/J}$ in accordance with Eq.(6). The two addends in brackets of Eq.(8) correspond to the conductance of the bonded (denoted by subscript "B") and separated (denote by subscript "S") sections of the graphene channel, respectively. Analytical expressions for the conductivities $\sigma_{B,S}$ are derived in **Appendix C**.

The conductivity of the bounded section has the form:

$$\sigma_B = \frac{2e^2}{\pi^{3/2}\hbar} \lambda_B \sqrt{n_S} . \qquad (9)$$

Here $e = 1.6 \times 10^{-19}$ C is elementary charge, $\hbar = 1.056 \times 10^{-34}$ J·s = $6.583 \times 10^{-16}$ eV·s is Plank constant, $v_F = 10^6$ m/s is characteristic electron velocity in graphene, $\lambda_B$ is mean free path in the graphene channel. The concentration of 2D electrons $n_S$ can be regarded constant voltage-independent value far from the FDWs, namely $n_S \approx |P_S/e|$ (see **Appendix C**). Using that for the most common case of electron scattering in graphene channel at ionized impurities in a substrate $\lambda_B[n_S] = \alpha\sqrt{n_S}$, where the proportionality coefficient $\alpha$ depends on the substrate material and graphene-ferroelectric interface chemistry [see eqs.(3.20-3.22) in Ref.[3] and **Appendix D**], we obtain from Eq.(9) the dependence $\sigma_B[n_S] = \frac{2e^2\alpha}{\pi^{3/2}\hbar} n_S \approx 8.75 \cdot 10^{-5} \alpha n_S$ (in Siemens). Taking into account that $P_S$ value can be 10 times smaller for thin films than its bulk value, the concentration vary in the range $n_S \cong (0.3 - 3) \times 10^{18}$ m$^{-2}$ depending on the film thickness, but should be regarded voltage- and coordinate- independent constant far from the FDW. Thus elementary estimates give $\sigma_B \cong (0.15 - 15) \times 10^{-3} \Omega^{-1}$ for reasonable ranges of $\lambda_B = (10 - 100)$ nm and $P_S = (0.05 - 0.5)$ C/m$^2$.

On the contrary, the main channel for electron scattering in the separated stretched section of structurally perfect graphene is collisions with acoustic phonons. In this case $\lambda_S(E) \sim 1/E$ [21]. This leads to paradoxical, however, well known result. Conductivity $\sigma_S$ doesn't depend on 2D electrons concentration in the graphene channel. Hence for further estimations we use a well known upper limit for $\sigma_S$ [3]:

$$\sigma_S = \frac{4e^2 \hbar \rho_m v_F^2 v_S^2}{\pi D_A^2 k_B T} \quad (10)$$

Here $\rho_m \approx 7.6 \cdot 10^{-7}$ kg/m$^2$ is 2D mass density of carriers in graphene, $v_S \approx 2.1 \cdot 10^4$ m/s is a sound velocity in graphene, Boltzmann constant $k_B = 1.38 \times 10^{-23}$ J/K; $D_A \approx 19$ eV is acoustic deformation potential that describes electron-phonon interaction. Expression (10) yields $\sigma_S \approx 3.4 \times 10^{-2} \Omega^{-1}$ at room temperature.

**Figures 4-6** present the conductance $G$ calculated from Eqs.(8)-(10) for different values of gate voltage $U$, electron mean free path in graphene $\lambda_B$, electron concentration $n_S$, binding energy $J$, graphene-ferroelectric separation $d$ and channel length $L$. The conductance in Eq.(8) is linearly proportional to the channel width $W$ that was chosen equal to 50 nm. Parameters $d_{33} \approx 10^3$ pm/V, $d_{31} \approx -450$ pm/V and Poisson ratio $v=0.3$ corresponds to PZT. Graphene Young's modulus is $Y = 1$ TPa and conductance $\sigma_S = 3.4 \times 10^{-2} \Omega^{-1}$. Other parameters vary within physically reasonable intervals, namely, concentration $0.1 \times 10^{18}$ m$^{-2} \leq n_S \leq 5 \times 10^{18}$ m$^{-2}$, binding energies $0.1$ J/m$^2 \leq J \leq 1$ J/m$^2$, electron mean free path 10 nm$\leq \lambda_B \leq$ 100 nm, channel length 50 nm $\leq L \leq 500$ nm and separation 0.1 nm $\leq d \leq 1$ nm. The behavior illustrated by **Figs.4-6** can be qualitatively explained by approximate expression (C.5) for the inverse conductance derived in **Appendix C**.

The conductance increases with $U$ increase; at that the increase is monotonic and faster than linear (i.e. "superlinear") at fixed other parameters [**Fig. 4**]. At that the increase is the most pronounced at small binding energies $J \leq 0.2$ J/m$^2$ [**Fig. 4(a)**], long mean free paths $\lambda_B \geq 50$ nm [**Fig. 4(b)**], small channel length $L \leq 100$ nm [**Fig. 4(c)**] and relatively high concentrations $n_S \geq 10^{18}$ m$^{-2}$ [**Fig. 4(d)**].

As one can see from **Figs.4** the conductance ratio $G(U)/G(0)$ doesn't exceed 1.25 for the case of graphene partial separation [presented by the red curve in **Fig.3(b)**] at realistic values of parameters. However, it can be essentially greater, if the domain stripe period $D$ is much shorter than the channel length $L$. For the case length $L$ is divided in two almost equal parts between the separated and bonded sections (i.e. $l \approx L/2$). If p-n-junctions at FDW don't change the general conductance of the graphene channel essentially [2] and the electron mean free pass $\lambda_{B.S} \ll D$ and $\sigma_S \gg \sigma_B$, Eq.(8) yields $\frac{G(U)}{G(0)} = \left[1 - \left(1 - \frac{\sigma_B}{\sigma_S}\right)\frac{l(U)}{L}\right]^{-1} \approx 2$. Thus the experimental situation [2] leads to rather high conductance ratio. Note that the ratio $G(U)/G(0)$ can be essentially greater than 2, e.g. in the case of mostly suspended graphene ($l \approx L$) with $\sigma_S \gg \sigma_B$. However, the possibility of such a limiting case needs special examination.

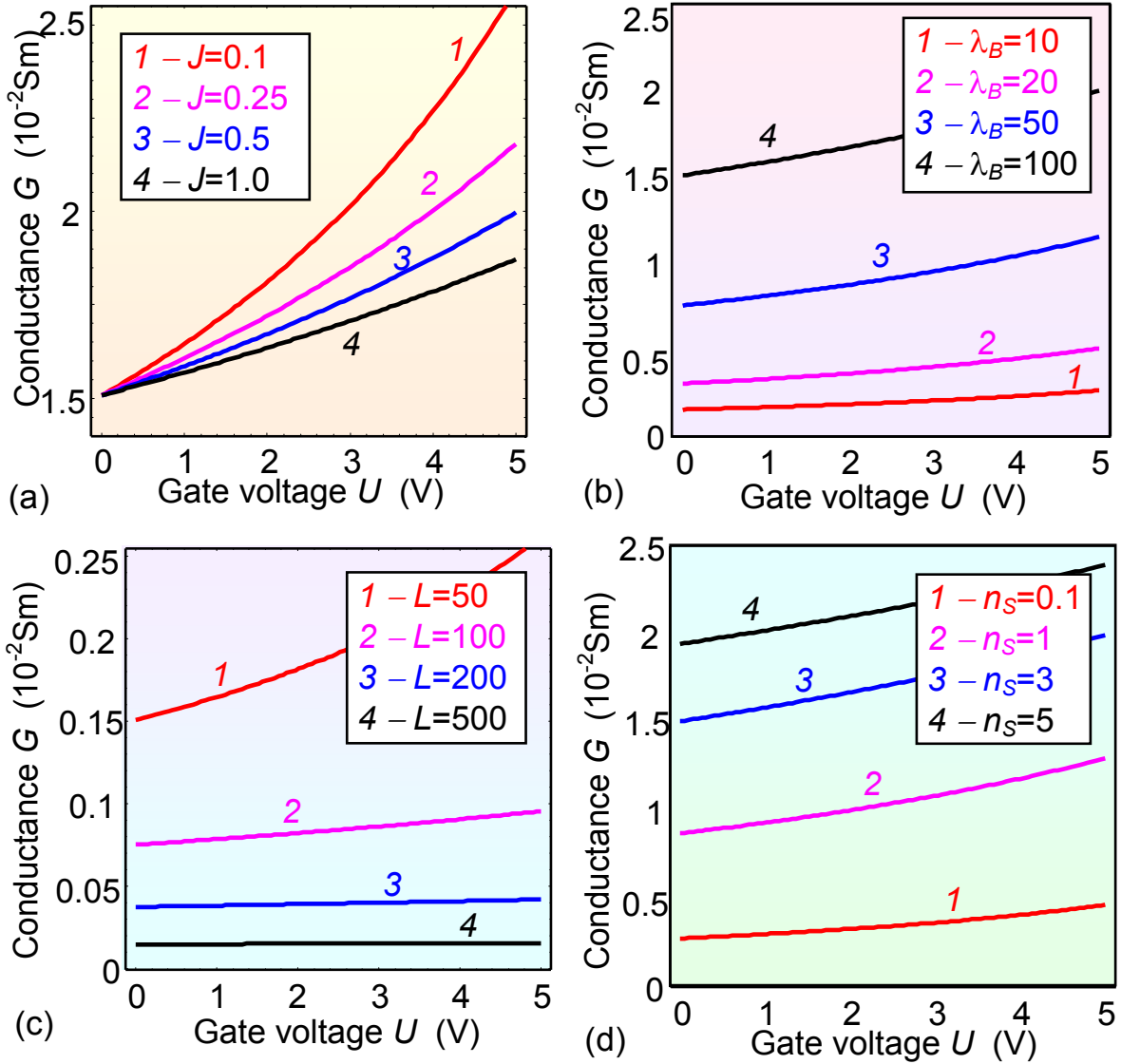

**FIG. 4.** Dependences of the conductance $G(U)$ on the gate voltage $U$ calculated for several values (curves 1 – 4) of binding energy $J$=0.1, 0.25, 0.5, 1.0 J/m² **(a)**; electron mean free path $\lambda_B$ = 10, 20, 50, 100 nm **(b)**; channel length $L$ = 50, 100, 200, 500 nm **(c)**; and concentration $n_S$ = (0.1, 1, 3, 5)×10¹⁸ m⁻² **(d)**. Piezoelectric coefficients $d_{33} \approx 10^3$ pm/V, $d_{31} \approx -450$ pm/V and Poisson ratio $v$=0.3 corresponds to PbZr$_{0.5}$Ti$_{0.5}$O$_3$, $n_S = 3 \times 10^{18}$ m$^{-2}$, $\lambda_B$=100 nm, $L$=50 nm, $W$= 50 nm, and separation $d = 0.5$ nm, graphene Young's modulus $Y = 1$ TPa, binding energy $J$=0.5 J/m$^2$.

Actually, the possibility $G(U)/G(0) \gg 1$ follows from the **Figs.5(a)-(c)**, which are colored contour maps of the conductance in dependence on coordinate pairs, such as $\{U,L\}$, $\{U,d\}$ and $\{U,J\}$. The horizontals lines in the plots **(b)** and **(c)** show the pronounced charges of $G(U)$ (from 0.15 Sm to 0.55 Sm) with $U$ increase from 0 to 5 V. Appeared that the conductance linearly increases with the ratio $|U|/L$ increase [see V-type colored contours in **Fig. 5(a)**]. The conductance dependences on the parameters $d$, $J$ and $|U|$ are in fact the dependence on the ratio $2|U|\sqrt[3]{Yd/2J}$, when its value becomes essential [compare colored contours in **Fig. 5(b)** and **5(c)**].

As seen from the color scale in **Fig.5(b)-(c)**, the pronounced changes of $G(U)$ (about 2.1 – 3.6 times) in comparison with its reference value $G(0) = W\sigma_B/L$ correspond to relatively high concentrations $n_S \geq 10^{18}$ m$^{-2}$, small binding energy $J \leq 0.2$ J/m$^2$, realistic separation $d = 0.5$ nm and small channel length $L \leq 100$ nm. Small binding energy makes the partial separation or complete exfoliation of graphene easier, and it separated areas – longer. Small channel length means the quasi-ballistic regime of current in the bonded graphene channel section. In the opposite case of a diffusive regime the total conductance is too low and cannot be changed essentially by a suspended section.

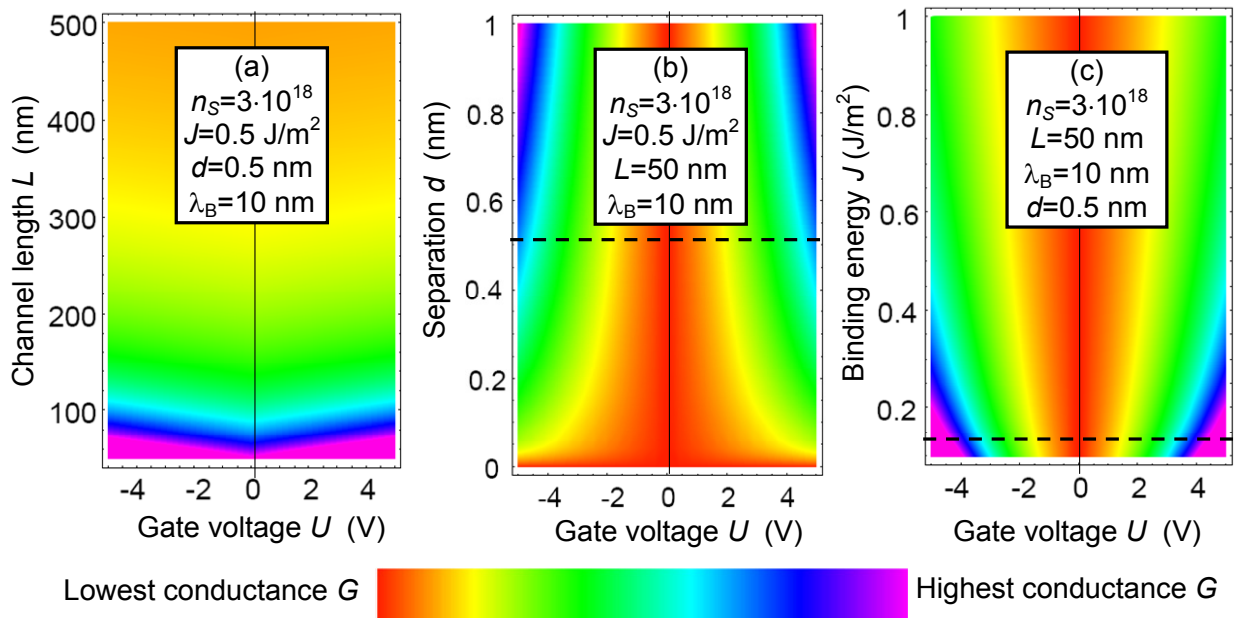

**FIG. 5.** Contour maps of the conductance $G(U)$ in different coordinates $\{U, L\}$ **(a)**, $\{U, d\}$ **(b)**, and $\{U, J\}$ **(c)**. Fixed parameters are listed in the legends. Other parameters are the same as in **Fig. 4.** Color scale ranges from 0.015 Sm (red) to 0.262 Sm (violet) for map **(a);** from 0.151 Sm (red) to 0.324 Sm (violet) for map **(b)**; and from 0.151 Sm (red) to 0.550 Sm (violet) for map **(c)**.

At fixed gate voltage (3V) the conductance strongly decreases with $L$ increase from 50 to 500 nm, at that the changes in one-two orders of magnitude correspond to small $\lambda_B \leq 50$ nm [**Fig. 6(a)**]. In fact we can interpret **Fig. 6(a)** as the demonstration of the extrinsic size effect of the conductance ("*L*-size effect"). Conductance dependence on the separation $d$ is a monotonically increasing one, but rather weak at $d > 0.2$ nm [**Fig. 6(b)**], while its dependence on $L$ and $\lambda_B$ is much more pronounced [compare to **Fig.6(a)**]. Conductance dependence on $J$ is a monotonically decreasing one, and also rather weak at $J > 0.2$ J/m$^2$ [**Fig. 6(c)**].

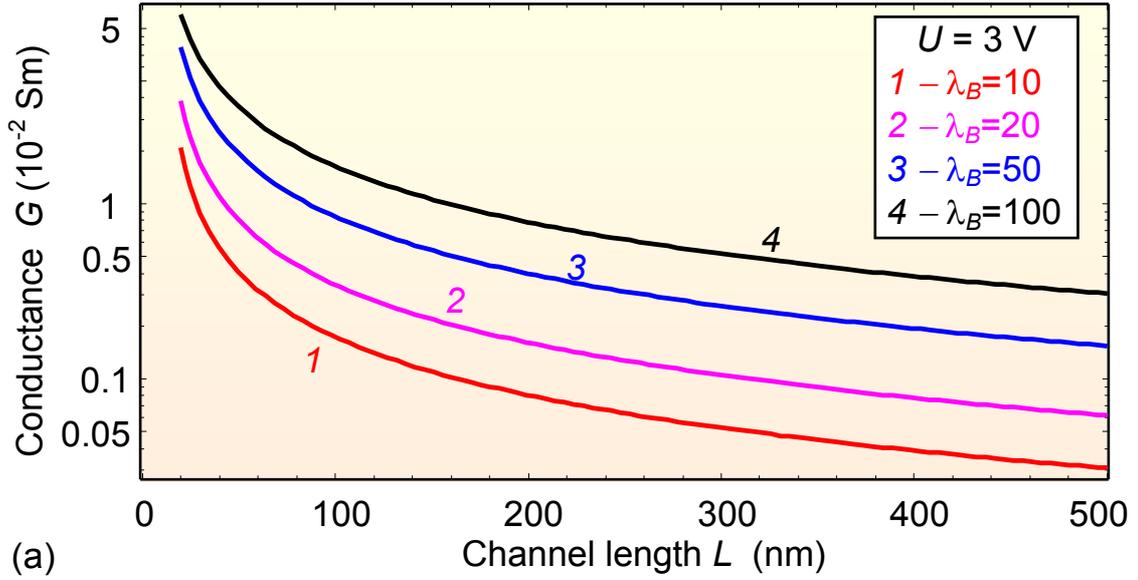
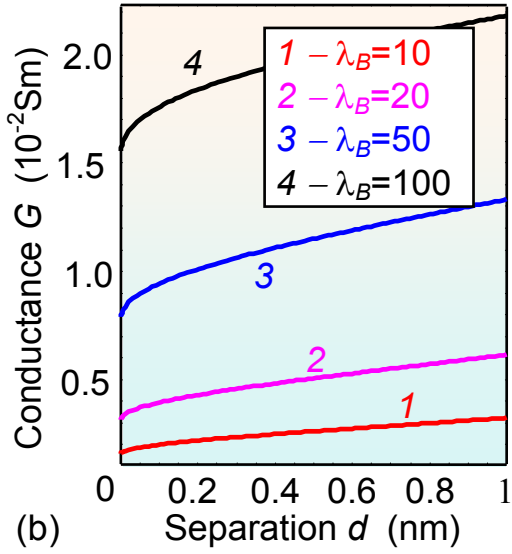
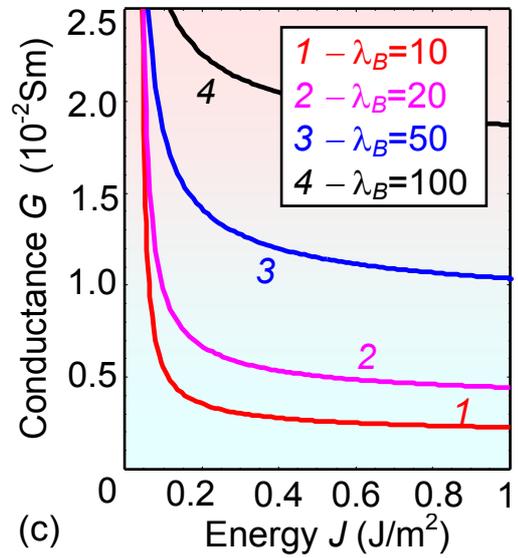

**FIG. 6. (a)** Dependences of the conductance $G(U)$ on the channel length $L$ calculated for gate voltage $U=3$V and several values of electron mean free path $\lambda_B = 10, 20, 50, 100$ nm (curves 1 – 4). **(b-c)** Conductance $G(U)$ in dependence on the separation $d$ **(c)** and binding energy $J$ **(d)**. Other parameters are the same as in **Fig. 4**.

The prediction of high ratios $G(U)/G(0)$ [see **Fig. 5(b)-(c)**] and its size effects [see **Fig. 6(a)**] is promising for advanced applications. Several possibilities are discussed in the next section.

# V. DISCUSSION OF POTENTIAL APPLICATIONS

Our calculations predict that the piezoelectric effect in the ferroelectric PbZr$_x$Ti$_{1-x}$O$_3$ (x≈0.5) leads to its surface displacement about (0.5 – 1) nm for the gate voltage (2 – 4)V and room temperature. The displacement can lead to the graphene separation from the substrate, since the density of the graphene-ferroelectric binding energy is relatively small and the Young's modulus of graphene is extremely high. The length of the separated section was estimated within a simple analytical model, showing that it can be 10 nm order or even longer. The separated sections of graphene channel induced by piezo-effect can cause interesting physical effects, which are interesting for fundamental physics. First, the conductance of graphene channel in diffusion regime changes essentially, because electrons in the stretched section scatter on acoustic phonons [21]. Second, mechanic vibrations of MHz range can be realised here [61]. Third, high pseudo-magnetic fields were reported for stretched graphene [62].

The consideration of the first of these three effects, performed in this work, are promising for advanced applications of graphene-on-ferroelectric with domain structure in *GFETs* and related *memory elements*, various *logic devices*, as well as for design of high efficient hybrid *electrical modulators, rectifiers* and *transducers of voltage-to-current type* [7-16, 21].

Abovementioned applications utilize that the concentration of 2D electrons in graphene channel on PZT substrate is extremely high (~10$^{18}$ m$^{-2}$). This value is at least in 1-2 orders higher than the maximal values obtained for graphene on ordinary mica substrate (see e.g.[63]). For dielectric substrate the concentrations higher than 10$^{17}$ m$^{-2}$ cannot be realized because of the electric breakdown in the substrate appearing under the gate voltage increase. Therefore the conductance of graphene on PZT substrate is one or two orders higher than for the case of a dielectric substrate. Also it linearly depends on 2D carrier concentration for the common case of dominant scattering at surface ionized impurities. However, as it was demonstrated above, the conductance strongly and nonlinearly depends on the gate voltage at small binding energies [see **Fig. 5(b)-(c)**] and channel length [see **Fig. 6(a)**].

The steep dependence $G(U)/G(0)$ induced by the piezo-effect in GFET can lead to the noticeable increase (from 1.5 to 3.5 times) of the voltage-to-current conversion coefficient in hybrid *electrical modulators*, which contain graphene-on-ferroelectric with FDWs. An essential advantage of the graphene conductance modulation by piezo-effect at FDWs proposed in this work in comparison with our earlier studies (see Ref.[16] and refs therein) is the following. The ferroelectric domains remain immobile in the horizontal direction *x* and relatively low gate voltage [64] creates almost instantly [65] the vertical z-displacement steps across the FDWs due to the piezoelectric effect [see **Figs. 7(a)-(c)**]. Since the dependence $G(U)$ is voltage symmetric, $G(U) = G(-U)$, the voltage-to-current conversion takes place at double frequency.

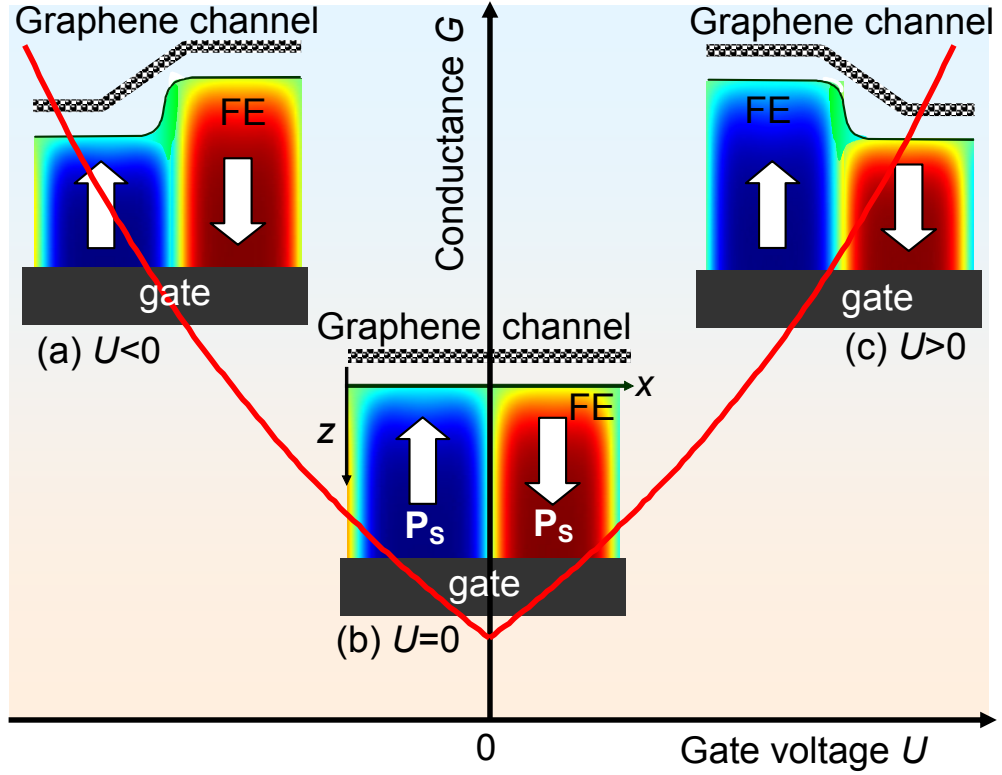

**FIG. 7.** Modulation of graphene channel conductance by piezo-effect in GFET on ferroelectric (FE) substrate. Main plot shows the schematic dependence of the channel conductance $G(U)$ on the gate voltage $U$. Insets **(a)** and **(c)** illustrate the vertical piezoelectric displacement of ferroelectric surface at FDW that causes partial separation of graphene sections induced by negative and positive gate voltages, respectively. The displacement and corresponding graphene separation are absent at $U$=0 [Inset **(b)**].

In contrast, the controllable motion of FDWs is required in horizontal x-direction along the channel for the noticeable changes of graphene conductance and electro-resistance without taking the piezoelectric effect into consideration [16]. The FDWs motion is sluggish and has a voltage threshold, i.e. requires gate voltages higher that the coercive voltage of polarization reversal. As a matter of fact it is much easier to create immobile FDWs, which are fixed spontaneously by existing pinning centers and lattice barriers, than to control the inertial motion of FDWs in the presence of unavoidable pinning centers [46]. Hence the electrical transducers, modulators and logic memory elements based on graphene-on-ferroelectric with FDWs, which operate using the piezoelectric displacement of immobile ferroelectric domain structure, can be significantly more efficient, much faster and better controllable than their analogs requiring the motion of FDWs. One can assume that due to the absence of the moving walls the durability of such modulators can be essentially longer. However the possibility of graphene complete exfoliation (instead of controllable partial separation) should be foreseen, estimated and excluded for real applications. Based on our theoretical estimates this can be realized by choosing optimal material of ferroelectric substrate and its thickness. The most serious challenge towards the applications is how to control the graphene-ferroelectric binding energy and separation.

The joint action of the piezoelectric and finite size effects should has a pronounced impact on the nonlinear hysteretic dynamics of the FDWs, stored charge and electro-resistance in the graphene-on-ferroelectric nanostructures with p-n junction potentials [16, 44]. This allows advanced applications of the nanostructures as ***ultra-sensitive piezo-resistive elements***. The idea of such element is to deposit the graphene cover on the surface of ferroelectric film with high piezoelectric response. Due to the strong dependence of graphene conductance and electro-resistance on the piezoelectric displacement of ferroelectric surface its electrical response to applied voltage and/or electromechanical efficiency can increase up to several times. The increase depends on the ferroelectric domain structure and FDWs amount, ferroelectric film thickness and its piezoelectric coefficients, graphene-ferroelectric separation and binding energy.

The obtained results indicate also on applications of a graphene sheet stretched by FDW in ***mechanical resonators*** and ***pseudo-magnetic field generators***, which could be a subject of further studies.

Finally, we'd like to underline that since graphene discovery and up to now the suspended areas in graphene channel were fabricated over trenches in the substrate (see e.g. [61]). Here we propose ***the alternative method of suspended areas fabrication***, based on the piezo-effect in a ferroelectric substrate. The method does not require any additional technological procedures like chemical etching or mechanical treating of the substrate surface. It can be used, among all, to fabricate electromechanical nanosystems for sensing of single molecules and even electrons (as it was proposed in [61]).

## VI. CONCLUSIONS

To resume, p-n junctions in graphene on FDWs have been actively studied recently, but the role of piezoelectric effect in a ferroelectric substrate was not considered. We propose a piezoelectric mechanism of conductance control in the GFET on a ferroelectric substrate with immobile domain walls. In particular we predict that the graphene channel conductance can be controlled by the gate voltage due to the piezoelectric elongation and contraction of ferroelectric domains with opposite polarization directions. At the same time the gate voltage does not change essentially the 2D carrier concentration in graphene. However it can create the bonded, separated, suspended and stretched sections of the graphene sheet, whose conductivity and resistivity are essentially different. Our calculations demonstrate the possibility of several times increase of GFET conductance for ferroelectric substrates with high piezoelectric response.

Taking into account that the conductance of the graphene-on-ferroelectric is essentially higher than the one of graphene on ordinary dielectric substrates, the predicted effect can be very useful for improvement and miniaturization of various electronic devices (such as advanced logic elements, memory cells, high efficient hybrid electrical modulators and voltage-to-current transducers with frequency doubling and relatively low operation voltages, and piezo-resistive elements).

Also we propose the alternative method of suspended graphene areas fabrication based on the piezo-effect in a ferroelectric substrate. The method does not require any additional technological procedures like chemical etching or mechanical treating of the substrate surface.

**Acknowledgements.** The publication contains the results of studies conducted by President's of Ukraine grant for competitive projects (grant number F74/25879) of the State Fund for Fundamental Research (A.I.K. and A.N.M.). A portion of this research was conducted at the Center for Nanophase Materials Sciences, which is a DOE Office of Science User Facility, CNMS2016-061.

## APPENDIX A

In particular case of perfect contact ($d = 0$) Eqs.(1)-(3) for the surface displacement simplify to:

$$u_3(x, d \to 0) = -\frac{2U}{\pi}\left[(d_{33} + (1+2\nu)d_{31})\arctan\left(\frac{x}{H}\right) + (d_{33} + \nu d_{31})\frac{x}{H}\ln\left(1 + \frac{H^2}{x^2}\right)\right]. \quad (A.1)$$

## APPENDIX B

It is natural to expect that the graphene separation occurs right to the point, where the normal component $F_n$ of the elastic tension force $F$ and binding force $F_b$ are equal [see **Fig.1(b)** and **Fig.3(b)**]. Namely

$$|F_n| = |F_b|. \quad (B.1)$$

Taking into account obvious expressions for the forces shown in **Fig 1(b)**,

$$|F| = \frac{YS\Delta l}{l}, \quad |F_n| = \frac{YS\Delta l}{l}\sin\alpha, \quad |F_S| = \frac{YS\Delta l}{l}\cos\alpha \approx \frac{YS\Delta l}{l}, \quad |F_b| \approx \frac{JS}{d}, \quad (B.2)$$

where $S$ is effective cross-section of carbon atom in graphene, and $l$ is the length of the separated graphene region. Because of the strong inequality $l \gg \Delta l$, we obtain that $h^2 + l^2 = (l + \Delta l)^2 \approx l^2 + 2l\Delta l$ and so $\Delta l = h^2/2l$. Substituting the approximations $\Delta l = h^2/2l$ and $\sin\alpha \approx \alpha \approx h/l$ in the expressions (B.2) we obtain the normal force $|F_n| \approx \frac{YSh^3}{l^3}$. Then substitution the expression (5) for $h_\infty < h$ in the equality (B.1), we finally get:

$$l = h \cdot \sqrt[3]{\frac{Yd}{2J}} > 2|U|(d_{33} + (1+2\nu)d_{31})\sqrt[3]{\frac{Yd}{2J}}. \quad (B.3)$$

The inequality in Eq.(B.3) originates from the inequality $h > h_\infty$.

## APPENDIX C

The conductance $G$ of graphene channel with length $L$ and width $W$, when some part of the channel with length $l(U)$ is suspended, can be presented according to Matiessen rule as [21]:

$$\frac{1}{G(U)} = \frac{1}{W}\left[\frac{L - l(U)}{\sigma_B} + \frac{l(U)}{\sigma_S}\right]. \quad (C.1)$$

The conductivities $\sigma_{B,S}$ can be presented as [21]:

$$\sigma_{B,S} = \frac{e^2}{\pi\hbar}\left(\frac{2E_F}{\pi\hbar v_F}\right)\lambda_{B,S}(E_F). \tag{C.2}$$

Here $e=1.6\times10^{-19}$ C is elementary charge, $\hbar = 1.056\times10^{-34}$ J·s = $6.583\times10^{-16}$ eV·s is Plank constant, $E_F \cong \hbar v_F \sqrt{\pi n_S(U,P_S)}$ is Fermi energy in graphene, $v_F = 10^6$ m/s is characteristic electron velocity in graphene, $\lambda_{B,S}(E_F)$ is electron mean free path in bonded and separated sections of graphene channel, respectively.

The dominant mechanism for electron scattering in graphene on substrate is scattering by ionised impurities in the substrate. In this case the electron mean free path $\lambda_B(E) \sim E$ [3]. Allowing for the well-known relation, $n_S(E_F) = \frac{E_F^2}{\pi\hbar^2 v_F^2}$, one leads to the following dependence of conductivity (C.2) on 2D electrons concentration $n_S$ and mean free path $\lambda_B$ in the graphene channel,

$$\sigma_B = \frac{2e^2}{\pi^{3/2}\hbar}\lambda_B\sqrt{n_S} \approx 8.75\cdot10^{-5}\lambda_B\sqrt{n_S} \text{ (in Siemens).} \tag{C.3}$$

The expression (C.3) should account for the dependence of $n_S$ on the ferroelectric polarization $P_S(x)$ and gate voltage $U$. Expressions used in Refs.[3, 66], show that the carrier density is proportional to the difference of the electric displacement normal components, $D_z(x,0) - D_z(x,-d)$, i.e. $en_S(x) \cong P_S(x) + \varepsilon_0\varepsilon_b E_z(x)$, where $\varepsilon_0 = 8.85\times10^{-12}$ F/m is a universal dielectric constant and $\varepsilon_b \cong 5$ is a background permittivity of PZT [67]. Since the spontaneous polarization saturates far from the FDW ($P_S(x) \to \pm P_S$), the concentration is

$$n_S \sim \left|\pm\frac{P_S}{e} + \frac{\varepsilon_0\varepsilon_b U}{e(H+d)}\right|. \tag{C.4}$$

At gate voltage $U = 1$V, film thickness $H = 50$ nm and separation $d = 0.5$ nm, the second contribution to the concentration in Eq.(C.4), $\frac{\varepsilon_0\varepsilon_b U}{H+d} \cong 0.001$ C/m$^2$, is much smaller than the fist one, $\frac{P_S}{e}$, for $P_S = 0.5$ C/m$^2$ corresponding to bulk PZT at room temperature.

The behavior illustrated by **Figs.4-6** can be explained by analytical expression (C.1) for the inverse conductance, where the length $l$ of the separated section is proportional to the ratio $l \sim |U|\sqrt[3]{Yd/2J}$ and the conductivity $\sigma_B \sim \lambda_B\sqrt{n_S} \sim n_S \sim \lambda_B^2$ according to Eq.(C.3). Thus the inverse conductance has two contributions from bonded and separated sections,

$$\frac{1}{G_B(U)} \sim \frac{1}{\lambda_B \sqrt{n_S}} \left( L - 2(d_{33} + (1+2\nu)d_{31}) \sqrt[3]{\frac{Yd}{2J}|U|} \right) \quad \text{and} \quad \frac{1}{G_S(U)} \sim \sqrt[3]{\frac{Yd}{2J}} \frac{|U|}{\sigma_S},$$

respectively. Since the inequalities $\sigma_S > \sigma_B$ or $\sigma_S \gg \sigma_B$ are valid in accordance with our estimates, the first contribution dominates if the strong inequality $l \ll L$ is valid, and so

$$G(U) \sim \frac{\lambda_B \sqrt{n_S}}{L - 2|U|(d_{33} + (1+2\nu)d_{31})\sqrt[3]{Yd/2J}}. \tag{C.5}$$

## APPENDIX D

Using eqs.(3.20-3.22) from Ref.[3] one get the expression for the relaxation time

$$\tau = \frac{\tau_0}{G[2r_s]} = \frac{\sqrt{n_s}}{2\sqrt{\pi}\, n_{imp}\, v_F\, G[2r_s]}, \tag{D.1}$$

$$\frac{G[x]}{x} = \frac{\pi}{4} + 3x - \frac{3\pi x^2}{2} + \frac{x(3x^2 - 2)\arccos[1/x]}{\sqrt{x^2 - 1}}. \tag{D.2}$$

Here $r_s$ is the interaction parameter (coupling constant), at that $r_s = 2.19/\varepsilon_b$ and $\varepsilon_b = 5$ is a background dielectric constant. The mean free path is given by expression:

$$\lambda_B = \frac{\pi}{2}\tau v_F = \frac{\pi}{2} v_F \frac{\sqrt{n_s}}{2\sqrt{\pi}\, n_{imp}\, v_F} \frac{1}{G[2r_s]} = \frac{\sqrt{\pi n_s}}{4 n_{imp}\, G[2r_s]} = \alpha \sqrt{n_s}, \tag{D.3}$$

For the coefficient α we finally get:

$$\alpha = \frac{\sqrt{\pi}}{4 n_{imp}\, G[2r_s]}\, [m^2], \tag{D.4}$$

where $G[2r_s]$ is a function of coupling constant and $n_{imp}$ is impurity concentration.